# A phase-locked frequency divide-by-3 optical parametric oscillator

A. Douillet, J.-J. Zondy, G. Santarelli, A. Makdissi, and A. Clairon

**Abstract-- Accurate phase-locked 3:1 division of an optical frequency was achieved, by using a continuous-wave (cw) doubly resonant optical parametric oscillator. A fractional frequency stability of $2\times10^{-17}$ of the division process has been achieved for 100 s integration time. The technique developed in this work can be generalized to the accurate phase and frequency control of any cw optical parametric oscillators.**

## I. Introduction

Novel compact optical frequency chains rely on visible to near IR broad frequency combs (spanning one octave) provided by a femtosecond laser that is spectrally broadened by self-phase modulation in photonic bandgap fibers [1-2]. The traceability of the frequency measurement to the microwave primary standard is provided by the phase-locking of the RF repetition rate of the femtosecond laser, which defines the comb mode spacing. Provided that we are able to synthesize a frequency difference equal to $\nu/N$ within the span of the comb ($\nu$ is the frequency to be determined and N an integer divider), the frequencies $\nu$, and the related subharmonics $\nu/N$ and $(1-N^{-1})\nu$ can be deduced in a straightforward step. Thus, the use of an optical division by N can considerably reduce the required span of the comb generator, and at the same time, reduce by $N^2$ the $\phi$ noise requirement for the phase-locking (or beatnote counting) of two of the frequency markers to the nearest comb modes.

Continuous-wave optical parametric oscillators (OPO) are promising optical frequency dividers. Divide-by-two (N=2) frequency degenerate OPOs have been demonstrated [3]. For a non-degenerate 3:1 OPO ($3\omega\rightarrow2\omega, \omega$), the idler wave must be frequency doubled so that the RF beat between the signal and the doubled-idler waves can be used to phase lock the device to zero or to an RF reference frequency [4]. The advantage of these OPO-based dividers is that they require only a single pump laser source to generate subharmonics outside the comb spectrum. In a free running OPO, the phase coherence of the subharmonic waves is ultimately limited by the phase diffusion noise stemming from the spontaneous parametric fluorescence. Once these waves are phase-locked, the coherent nature of the parametric division process leads to subharmonic phase-noise reduction by $N^2$ relative to the pump laser phase noise. We report here on an accurate phase-locked optical frequency divider of a diode laser operating in the range 840-850 nm. We show that even with a weak beatnote signal-to-noise ratio, and without a fast electro-optic cavity length actuator, it is possible to achieve a very high fractional stability and long-term operation of the divider.

Laboratoire Primaire du Temps et des Fréquences (BNM-LPTF), Bureau National de Métrologie / Observatoire de Paris, 61, Avenue de l'observatoire, F-75014, Paris (France)

## II. The DRO Divider Setup

### A. Experimental

The doubly resonant optical parametric oscillator (DRO) is pumped by a master-oscillator power amplifier (MOPA) AlGaAs diode laser which is optically injected by a AlGaAs extended-cavity diode laser (Fig.1). The pump wavelength is $\lambda_p$=843.06 nm and its short-term linewidth 100 kHz. The available power at the DRO input is 0.4 W. The nearly spherical DRO cavity consists of two highly reflective (at the signal and idler wavelengths) ZnSe mirrors spaced by L=106 mm. Their radius of curvature is R=50 mm. The cavity length can be tuned using a 20 mm long piezoelectric transducer (PZT). A multi-grating, 19 mm long periodically poled lithium niobate (PPLN, poling period $\Lambda$=29.2 µm) is temperature phase-matched (T= 100°C) for the $3\omega\rightarrow(2\omega\pm\delta)+(\omega\mp\delta)$ interaction. The $\delta$ quantity ($\delta<<\omega$) represents the small radio-frequency mismatch from perfect division ratio. An intra-cavity $CaF_2$ Brewster plate was inserted in order to couple out as much as 3 mW of idler power (without the plate only 200 µW exits the rear mirror), at the expense of an increased oscillation threshold (65 mW compared with 15 mW). The available signal power is 6 mW. We perform the second-harmonic generation (SHG) of the idler wave in a PPLN sample ($\Lambda$=35 µm, T=68°C). About 3 mW of signal ($2\omega\pm\delta$) and 5 nW of useful doubled-idler ($2\omega\mp2\delta$) waves are mixed on a 5 GHz bandwidth avalanche photodiode. The resulting $3\delta$ frequency beat signal is used to control the division ratio of the DRO. Further details of the experimental setup can be found elsewhere [5].

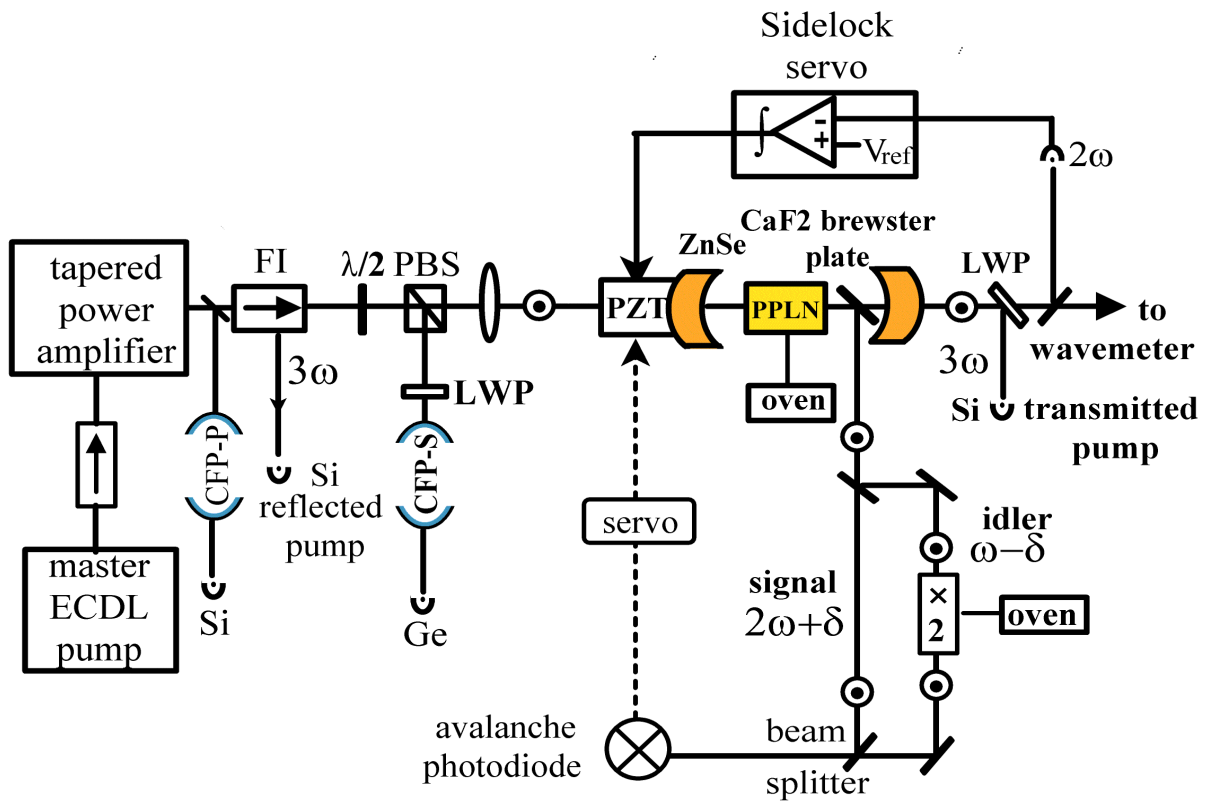

Fig. 1. The DRO, idler SHG and 3δ beanote detection set-up. FI: Faraday isolator; PBS: polarizing beamsplitter; LWP: long-wave pass filter; CFP: confocal Fabry-Perot; PZT: piezo-electric transducer.. Two wavemeters (not shown) enable to tune the DRO close to the 3:1 division.

### B. *Single mode pair operation and beatnote detection*

Due to the weak dispersion of the extraordinary index of refraction of lithium niobate from 2.53 µm to 1.26 µm, the signal and idler free spectral ranges (FSR ≈ 1.15 GHz) differ by only 1 or 2 MHz. Consequently more than 700 nearly resonant signal-idler mode pairs can experience similar gain within a mode cluster, giving rise to permanent axial mode hops (a few gigahertz away) or cluster hops (a few tens of gigahertz) under free running DRO operation. The DRO natural tendency to mode-hop requires a permanent and effective sub-nanometric servo control of its cavity length in order to detect a stable beatnote prior to the phase-locking step. Single mode pair operation is achieved with a standard side-of-fringe (*sidelock*) servo. The error signal of this servo loop is built by comparing the signal wave power detected by an InAs photodiode to a reference voltage that sets the frequency detuning $\Delta_s = \omega_s - \omega_c^s$ of the signal wave ( $\omega_s$ ) from the cavity eigenmode frequency $\omega_c^s = qc/2L_s$ ( $L_s$ is the signal wave optical path length of the resonator and $q$ the mode number). This error signal is then integrated and fed back to the PZT transducer. The bandwidth of the sidelock integral servo is limited to 1 kHz by mode-hops. Let us note that this sidelock servo is not a pure frequency servo. While it maintains constant both relative detunings, it does not correct for the drift of the cavity eigenmode frequencies. Furthermore it converts any power fluctuation into a frequency correction. The resulting beat at 3δ under pure sidelock servo is hence expected to be noisy. Under sidelock servo, the DRO can oscillate on a single mode pair during 5-15 minutes, which leaves enough time to proceed to the phase-locking step.

The selection of the appropriate mode pairs (e.g. those yielding a beanote falling within the 5 GHz electronic detection bandwidth) is the first experimental difficulty we had to overcome. Because of the high mode pair density, the particular pair captured by the sidelock is usually random. Fine tuning of the OPO-PPLN temperature and of the sidelock reference voltage have to be repetitively processed until a 3δ beat signal in the frequency range 200 MHz-5 GHz, with a typical power level of –50dBm and a signal to noise (S/N) ration of ~30 dB in a 100 kHz bandwidth, is detected on the spectrum analyzer. The beat frequency drift is rather low (less than 1 MHz per minute). Unfortunately the large short-term frequency jitter (~300 kHz peak-to-peak excursion) under sidelock servo combined with the limited bandwidth of the unique PZT transducer (<10 kHz) prevented us from performing a direct phase locking, by summing (after a proper filtering) the phase error derived from a stable RF source with the sidelock error signal. With a wide enough control bandwidth provided for instance by an electro-optic phase modulator [3-4], such a procedure would readily work. In the present case, since the sidelock servo (which maintains the pre-requisite single mode operation) has to be maintained during the phase locking step, various nested servo loops that we shall detail in the following section have to be implemented simultaneously in order to achieve a phase locked divider.

## III. THE PHASE-COHERENT FREQUENCY DIVIDER

The ways we found to circumvent these difficulties are multiple. Given the noisy feature of the free running beatnote, the procedure must comply to the following preliminary steps. At first we reduced the large beat frequency fluctuations by replacing (in a continuous way to prevent a mode hop) the sidelock servo by a much quieter *pure* frequency lock servo. This consequently reduces the required dynamical range of the phase servo by minimizing the beat frequency jitter and drift. Secondly we increased further the dynamic range of the phase error signal by dividing the beanote frequency (by 128) in order to reduce the control bandwidth required to achieve a stable (free of phase jumps) phase locked loop. Finally, to increase the long-term operation of the phase-locked divider (limited by the occurrence of a mode hop), we developed a method allowing the coexistence of the frequency (FL) and the phase locked (PL) loops. Fig.2 sketches the various nested loops that we have implemented to achieve a long-term phase locked, mode-hop-free operation during more than one hour.

Let us describe now in detail the phase locking steps. Once a “free running” beatnote is captured according to the procedure described in section II.B, its frequency is down-converted to a fixed RF frequency of 100 MHz by mixing it with a microwave synthesizer synchronized to a high stability 10 MHz reference from an H-maser. To avoid false triggering of the digital by-128 frequency divider, due to additive AM noise, we have improved the S/N ratio of the down-converted beatnote by phase-locking a 100 MHz voltage-controlled oscillator (VCO) to it. This servo loop is denoted the *tracking loop* in fig.2. The bandwidth of this phase loop is wide enough (1 MHz) to track almost perfectly the phase/frequency fluctuations of the optical beat.

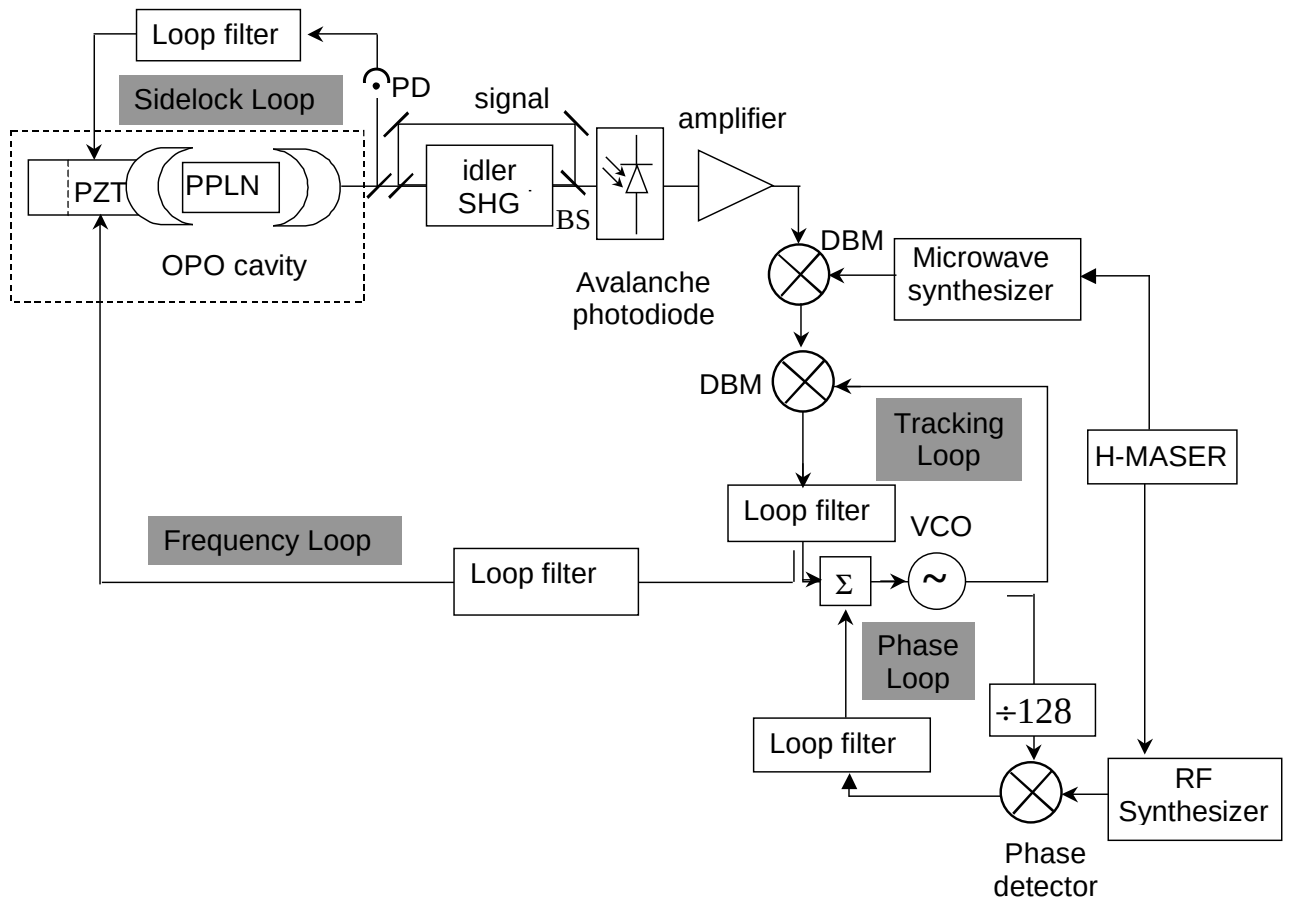

Fig. 2. The phase/frequency nested loops block diagram, comprising. The sidelock loop is disabled once the frequency loop is effective. DBM: double-balanced mixer; PD: InAs photodiode.

The 20 dB S/N ratio achieved in the loop bandwidth makes very improbable noise-induced cycle slips. The higher S/N ratio at the VCO output also allows convenient counting of the beat frequency with a RF counter referenced to the H-maser, for further diagnosis of the stability of the DRO divider.

When the tracking loop is closed, the correction voltage applied to the VCO varicap is proportional to the frequency difference between the down-converted instantaneous frequency and the free-running VCO frequency. This voltage provides the error signal of the frequency loop (FL) of fig.2. By feeding back (with an appropriate sign) this correction voltage to the PZT, one slaves in turn the jittered 3δ beat frequency to the natural VCO frequency. Tuning the VCO frequency results hence in the tuning of the 3δ beat frequency. To insure a permanent and stable control of the DRO cavity the FL and sidelock error signals are summed before integration and both loops are used simultaneously for a while. The stable coexistence of both loops is possible because both error signals have a large dynamic range and provide similar physical information. When the gain of the FL servo is sufficiently large the sidelock loop can be disabled without inducing a mode hop. Operating the divider under pure FL provides a strong reduction of the beatnote fluctuations down to a few kiloHertz level compared with 300 kHz when both loops are coexisting, meaning that the sidelock acts as a noise source for the FL. As a consequence, under pure FL operation the DRO tendency to mode hop is then notably reduced. In Fig. 3 we have plotted the Allan standard deviation $\sigma_y(\tau)$ of the DRO, under sidelock (upper curve) and frequency lock (lower curve). The data were derived from 3δ frequency measurements using 1s counting gate time. The quantity $\sigma_y$ , where y is defined as $y=\langle\delta(t)\rangle/\nu_i$ and $\nu_i$=118.6 THz is the idler frequency, measures the stability of the optical frequency division. Fig 3 illustrates the additional frequency noise brought by the sidelock servo, as compared with the FL servo. The origin of this additional noise stems from the hybrid nature of the information carried by the side-of-fringe discriminator signal, as discussed in section II. In particular we suspect a specific thermally induced AM to FM noise conversion processes [5] to be the cause of the large frequency jitter of the 3δ beat under pure sidelock servo. Indeed, the 300 kHz jitter cannot be reasonably imputed either to the small pump laser amplitude noise (1% intensity fluctuation would correspond to 10 kHz frequency excursion given an estimated cavity linewidth of ~5 MHz) or to the acoustical cavity perturbations since they are efficiently corrected by the FL loop.

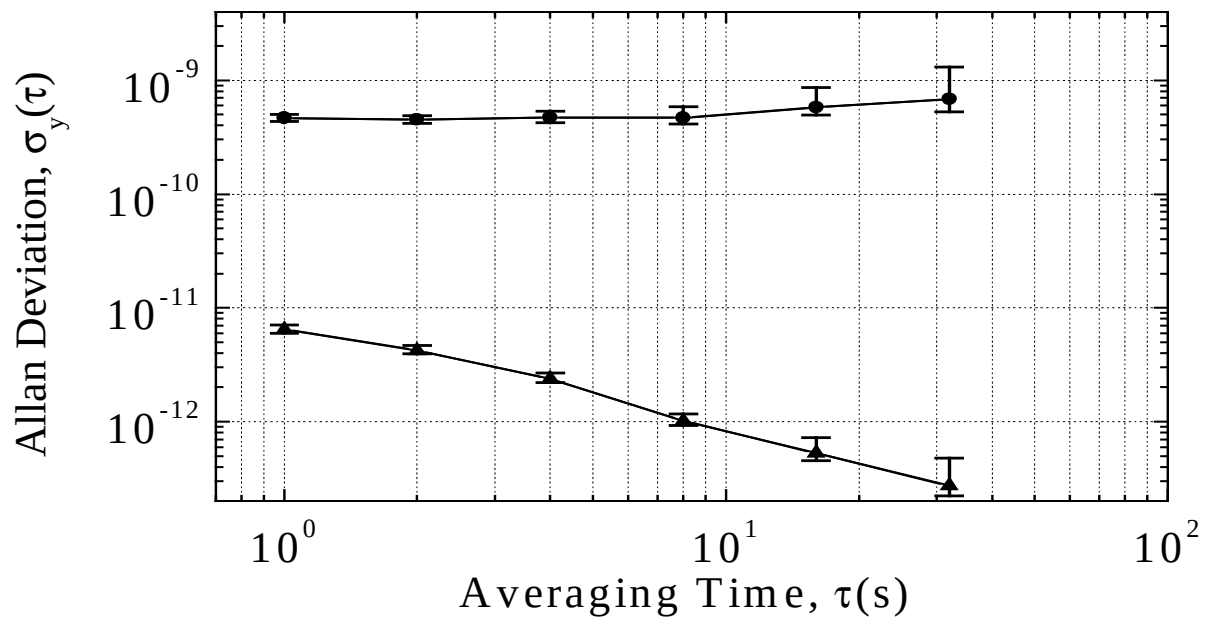

Fig. 3. Allan standard deviation of the 3:1 divider frequency noise, relative to the idler frequency under sidelock (circles) and frequency lock (triangles). The frequency locking of the OPO beatnote to the VCO decreases the beat noise by almost two decades at τ=1s.

To this point, the beat fluctuations are low enough to allow the final phase-locking step. The 100 MHz VCO output is digitally divided by 128 and the resulting output signal mixed in a digital phase detector with a 781 kHz synthesizer, synchronized to the H-maser. The output phase error signal has a 128π peak-to-peak dynamic range, because of the division stage, and hence allows a reduced phase locking loop bandwidth. The most natural way to control the beatnote frequency when the FL loop is closed is to change the VCO varicap voltage. We took advantage of this entry point to phase lock the DRO divider. In this way we achieved in fact a cascaded phase locking of the beatnote. This loop is denoted the *phase loop* (PL) in fig.2. Hence the VCO itself acts as a second fast transducer of the DRO divider. Furthermore, summing on the VCO varicap the tracking and phase lock correction signals leads to a simultaneous operation of both the FL and the PL loops. Under FL and PL action, the down-converted beat frequency fluctuations dramatically dropped to the millihertz level for τ=100 s counting gate time.

The residual phase error at the output of the phase detector was displayed on an FFT spectrum analyzer (fig.4) and used to adjust the relative gains of the FL and PL by minimizing the residual phase error variance $\sigma_\phi^2 = \int S_\phi(f)df$ . The small peak at 22 kHz is due to a PZT resonance frequency which ultimately limits the overall gain of the system. The sharp low-frequency peaks at ~ 50 Hz and ~100Hz are due to the electrical power supply pickups and the remaining broader peaks (20, 250 and 1000 Hz) probably due to the acoustical resonance of the DRO mechanical structure. The major contribution to $\sigma_\phi^2$ arises from the broad noise feature between $10^3$ and $10^4$Hz in fig.4, which we identified as possibly originating from thermo-optically induced AM to FM noise conversion (a weak thermal self-stabilization of the optical path length of the DRO was observed as in ref. [5]). The total phase noise variance deduced from the spectrum of fig.4 is $\sigma_\phi^2 = 1.2$ rad$^2$ corresponding to only $\exp(-\sigma_\phi^2) = 30\%$ of the beatnote energy in a narrow coherent peak. Indeed, the measured residual 3δ beatnote linewidth was 3 kHz, and no coherent peak was clearly visible. The DRO could remain phase-locked for more than one hour, without any mode hop, which highlights the robustness of the combined FL/PL loops. The numerical modeling of the action of the nested loops reveals that such a coexisting FL/PL loop system can tolerate very large frequency jumps (hence the reduced mode hop events). In particular, when the phase goes out of lock the FL brings back the frequency error within the capture range of the phase lock loop, shortening its recapture delay by a factor of 100.

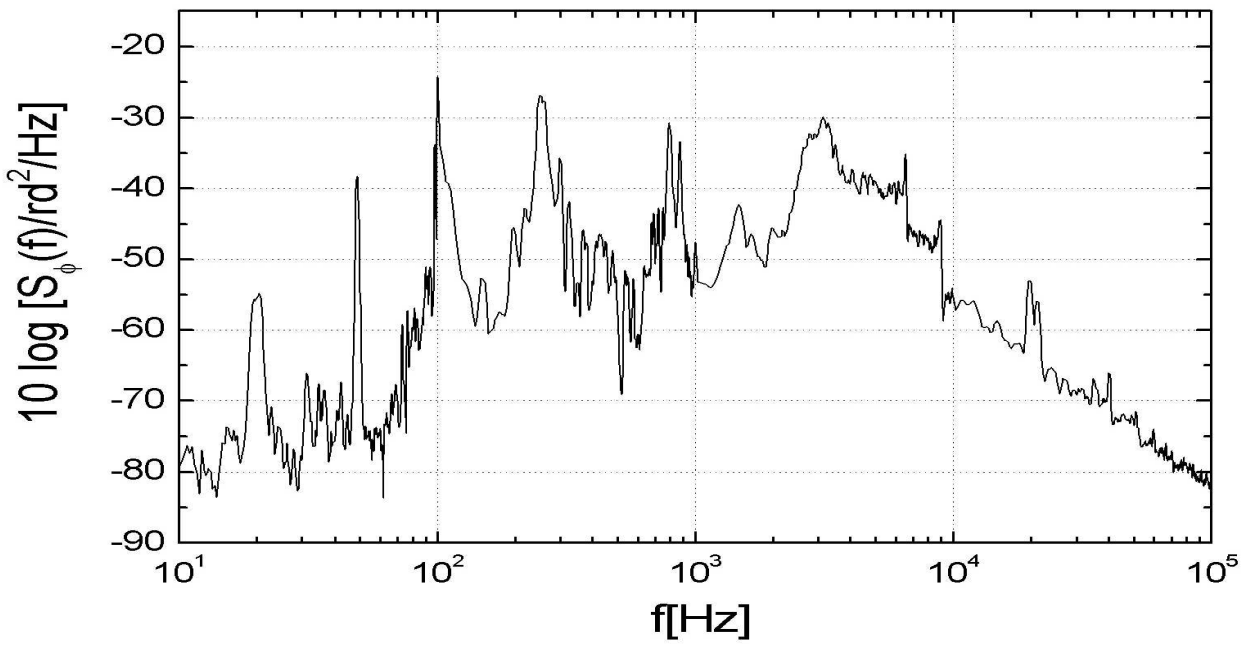

Fig. 4. Residual phase noise spectrum of the phase locked divider (relative to the offset δ quantity scaled to the idler frequency), measured from the phase error signal, when all loops in fig.2 (except the sidelock loop) are closed.

The Allan standard deviation of the phase-locked divider is reported in fig.5. The improvement of the short-term ($\tau$=1s) stability of the division process is nearly three orders of magnitude compared with the stability under FL action only (see fig.3). For longer integration time (up to $\tau$=100s gate time), the stability decreases as $\tau^{-1}$, meaning that the residual phase excursion is bounded to less than $2\pi$ in the long term. This also demonstrates that no cycle slips occurred during the measurement. This is a clear signature of the phase locking of the nearly subharmonic waves.

## IV. Conclusion

We have achieved a phase-coherent optical frequency division by 3 of 843 nm light, using a doubly resonant optical parametric oscillator. A relative optical divider frequency stability of $2\times10^{-17}$ for 100 *s* integration time, at the idler wave is achieved, which highlights the potential resolution of parametric dividers in a frequency measurement setup. This PPLN-DRO also generates a comb of phase-locked frequencies ($\omega_p/3$, $2\omega_p/3$, $\omega_p$, $4/3\omega_p$, $5\omega_p/3$, $2\omega_p$) from various non-phase matched conversion mixings that occur within the chip. Measuring any interval between these sidebands leads to a simple determination of all frequencies of the comb. Combined with a femtosecond Ti:Sa laser clock [2], such a phase-locked OPO subharmonic comb can provide a direct microwave to deep mid-IR link or reduce by one third the required span of repetition rate stabilized femtosecond lasers. An alternative implementation of 3:1 DRO dividers would consist in patterning both OPO and idler SHG interactions on a single PPLN chip. The intra-cavity cascaded nonlinearities would then induce a self phase locking of the signal and idler waves by mutual injection locking. Such an all-optical phase locking would avoid the use of the complex electronic phase-lock loops.

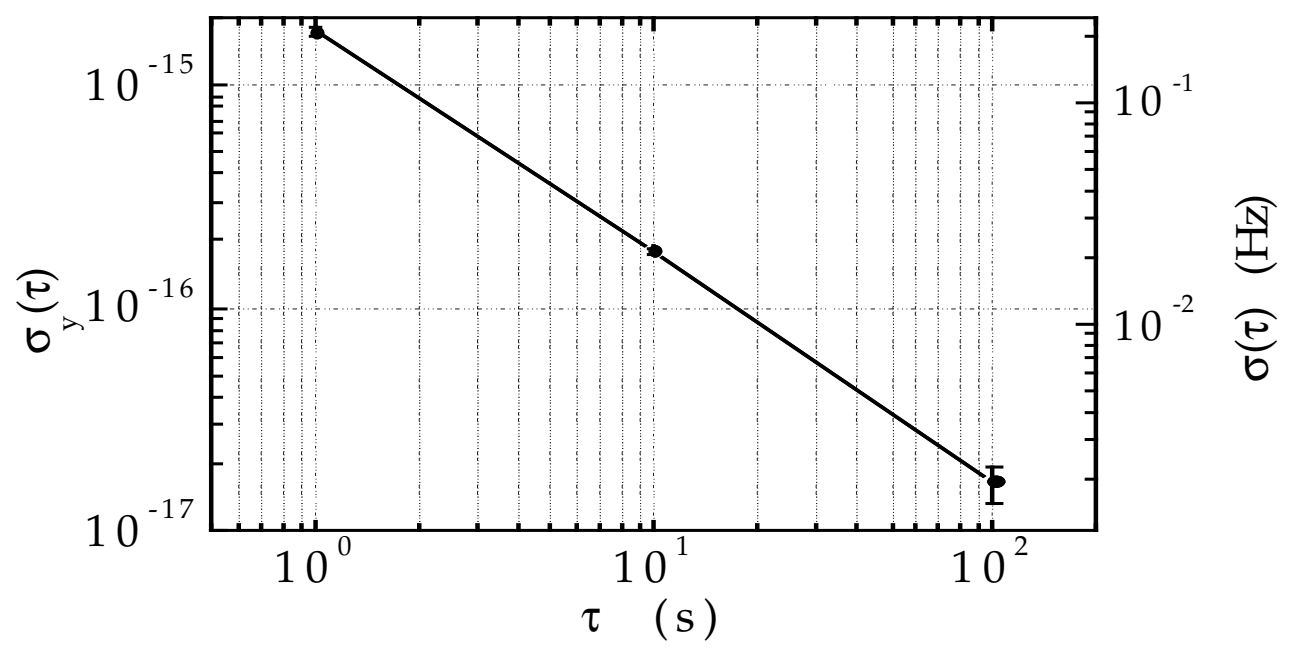

Fig. 5. Allan standard deviation of the 3:1 divider frequency noise, relative to the idler frequency (left axis) and absolute variance of δ, as a function of the integration counter gate time ($\tau$ =1, 10, 100s).